\documentclass[prl,  twocolumn, superscriptaddress, longbibliography]{revtex4-2}
\def\be{\begin{equation}}
\def\ee{\end{equation}}
\def\bi{\begin{itemize}}
\def\ei{\end{itemize}}
\usepackage{braket}

\usepackage{xcolor}
\usepackage{graphicx}
\usepackage{amsmath,amstext,amssymb,bm,color,times}
\usepackage[english]{babel}
\usepackage[T1]{fontenc}
\usepackage[utf8]{inputenc}
\usepackage[colorlinks=true,citecolor=blue,linkcolor=magenta]{hyperref}
\usepackage{lipsum}

\usepackage{pdfpages}
\usepackage{pgffor}

\makeatletter
\AtBeginDocument{\let\LS@rot\@undefined}
\makeatother

\begin{document}

%%%%%%%%%%%%%%%%%%%%%%%%%%%%%%%%%%%%%%%%%%%%%%%%%%%%%%%%%%%%%%%%%%%%%%%%%%

\title{Quantum phase transition dynamics in the two-dimensional transverse-field Ising model}

%%%%%%%%%%%%%%%%%%%%%%%%%%%%%%%%%%%%%%%%%%%%%%%%%%%%%%%%%%%%%%%%%%%%%%%%%%
\author{Markus Schmitt}
\affiliation{Institute for Theoretical Physics, University of Cologne, 50937 Cologne, Germany}
\author{Marek M. Rams}
\affiliation{Jagiellonian University, Institute of Theoretical Physics, {\L}ojasiewicza 11, PL-30348 Krak\'ow, Poland}
\author{Jacek Dziarmaga}
\affiliation{Jagiellonian University, Institute of Theoretical Physics, {\L}ojasiewicza 11, PL-30348 Krak\'ow, Poland}
\author{Markus Heyl}
\affiliation{Max Planck Institute for the Physics of Complex Systems, N{\"o}thnitzer Stra{\ss}e 38, Dresden 01187, Germany}
\affiliation{Theoretical Physics III, Center for Electronic Correlations and Magnetism, Institute of Physics, University of Augsburg, D-86135 Augsburg, Germany}
\author{Wojciech H. Zurek}
\affiliation{Theory Division, Los Alamos National Laboratory, Los Alamos, New Mexico 87545, USA}

\date{\today}
%%%%%%%%%%%%%%%%%%%%%%%%%%%%%%%%%%%%%%%%%%%%%%%%%%%%%%%%%%%%%%%%%%%%%%%%%%

\begin{abstract}
The quantum Kibble-Zurek mechanism (QKZM) predicts universal dynamical behavior near the quantum phase transitions (QPTs).
It is now well understood for the one-dimensional quantum matter.
Higher-dimensional systems, however, remain a challenge, complicated by the fundamentally different character of the associated QPTs and their underlying conformal field theories.
In this work, we take the first steps toward theoretical exploration of the QKZM in two dimensions for interacting quantum matter.
We study the dynamical crossing of the QPT in the paradigmatic Ising model by a joint effort of modern state-of-the-art numerical methods, including artificial neural networks and tensor networks.
As a central result, we quantify universal QKZM behavior close to the QPT. We also note that, upon traversing further into the ferromagnetic regime, deviations from the QKZM prediction appear. We explain the observed behavior by proposing an {\it extended QKZM} taking into account spectral information as well as phase ordering.
Our work provides a testing platform for higher-dimensional quantum simulators.
\end{abstract}
\maketitle

%%%%%%%%%%%%%%%%%%%%%%%%%%%%%%%%%%%%%%%%%%%%%%%%%%%%%%%%%%%%%%%%%%%%%%%%%%%%%%

%%%%%%%%%%%%%%%%%%%%%%%%%%%%%%%%%%%%%%%%%%%%%%%%%%%%%%%%%%%%%%%%%%%%%%%%%%%%%%
\section{Introduction}
%\label{sec:intro}
%%%%%%%%%%%%%%%%%%%%%%%%%%%%%%%%%%%%%%%%%%%%%%%%%%%%%%%%%%%%%%%%%%%%%%%%%%%%%%
The near-critical region of continuous symmetry breaking phase transition is characterized by critical slowing down reflected in an asymptotically divergent relaxation time scale and correlation length. When a many-body system is quenched -- driven across the critical point at a fixed rate -- critical slowing down will prevent its order parameter from keeping up with what would have been the instantaneous equilibrium. In particular, the correlation length -- hence, the size of the fluctuating domains within which the order parameter is approximately uniform -- will lag behind the state implied by the externally imposed conditions. Thus, instead of an infinite range order predicted in equilibrium for the post-transition broken symmetry phase, one is left with a mosaic of fluctuating domains. Independent choices of disparate broken symmetry states within each domain result in excitations and lead to the formation of stable topological defects.

Inevitable appearance of  the topological defects in the wake of the cosmological phase transitions was pointed out by Kibble \cite{K-a, *K-b, *K-c}, who  tied their emergence and stability to the homotopy group, and suggested that their post-transition density will be set by thermal activation (which would imply that their density is independent of the rate of quench). Kibble also noted that, in the wake of the Big Bang, the Hubble radius at the epoch of the transition will imply a lower bound on defect density and emphasized that -- even at that density -- they can have dramatic consequences for the subsequent evolution of the Universe. 

The role of critical slowing down in the creation of topological defects in laboratory phase transition was pointed out by one of us \cite{Z-a, *Z-b, *Z-c}. The key difference with the cosmological setting is that now the relativistic causality (reflected in the Hubble radius or the light cone) does not play any useful role. Rather, it is the combination of the critical slowing down and the limits imposed on the growth of the correlation length that decide the size of domains that can independently select how to break symmetry. Hence, one can infer properties of the post-transition broken symmetry state arising from this non-equilibrium process -- including the dependence of the density of topological defects and other excitations deposited by the quench on the speed of the transition -- from the universal equilibrium scalings and the quench rate 
\cite{Z-a, *Z-b, *Z-c, 
Z-d,
 KZnum-a,
 KZnum-b,
 KZnum-d,
 KZnum-g,
 KZnum-j}. 
This leads to the Kibble-Zurek mechanism (KZM) that is still being tested both numerically  and in laboratory experiments 
~\cite{ KZexp-c,
KZexp-d,
KZexp-e,
KZexp-f,
% KZexp-g,
KZexp-gg,
KZexp-h,
% KZexp-j,
KZexp-k,
KZexp-l,
KZexp-m,
KZexp-n,
KZexp-o,
KZexp-p,
KZexp-q,
KZexp-r,
KZexp-s,
KZexp-t,
KZexp-v,
KZexp-x}.

The KZM applies  not only to thermal phase transitions but has also been extended to the case of quantum critical points of local Hamiltonians~\cite{QKZ1,
QKZ2,
QKZ3,
d2005,
d2010-a,
d2010-b,
QKZteor-b,
QKZteor-e,
QKZteor-f,
QKZteor-h,
QKZteor-j,
QKZteor-k,
QKZteor-m,
KZLR3,
QKZteor-r}.
By now, the corresponding quantum KZM (QKZM) has been extensively explored and understood for one-dimensional (1D) models, where very recently pioneering experiments in systems of Rydberg atoms have confirmed its predictions in interacting quantum matter~\cite{Lukin18}. Other experiments have already started to explore QKZM~\cite{QKZexp-e, QKZexp-f, QKZexp-g,deMarco2}, considered effectively mean-field type systems~\cite{QKZexp-a, QKZexp-b, QKZexp-c,QKZexp-d}, or spin models immersed in environment~\cite{adolfodwave,2dkzdwave}.

Going beyond 1D for interacting quantum matter appears central not only because of the real-world relevance of 2D systems but also because of fundamental differences compared to 1D.
Specifically, conformal field theories describing the universal properties at 2D quantum phase transitions are fundamentally different from their 1D counterparts in that they are generally of a strongly interacting nature. This leads to a  change of the character of excitations and the dynamics in the vicinity of the critical point as well as subsequent evolution.
While some results on nonequilibrium real-time evolution are available, see~\cite{QKZ3,Sengupta2D,MondalKitaev,KZnum-m}, the theoretical and numerical treatment of interacting quantum 2D systems poses severe challenges. The question to what extent the QKZM also applies to interacting 2D quantum many-body systems has so far remained largely open.

Very recent developments in the Rydberg atom quantum simulator platforms~\cite{rydberg2d1,rydberg2d2,Semeghini2021} or superconducting qubits~\cite{Satzinger2021} have nevertheless opened the way to access quantum dynamics in 2D at large scales and at a high level of control with the potential to target the outstanding challenge of QKZM in 2D~\cite{rydberg2d1}.
A particularly notable step forward is a very recent experiment in a Rydberg atom array~\cite{rydberg2d1}. In the vicinity of the critical point it is consistent with the QKZM scaling of the correlation length obtained through a fitting procedure to the post-quench correlation function extracted from a $16\times 16$ square lattice. 

In this work, we provide a large-scale numerical analysis of the dynamics in the 2D transverse-field Ising model in systems of various sizes with open and periodic boundary conditions. Going beyond the previous analysis of experimental data, we access the full scaling form of the correlation function, which has the advantage that no prior assumptions on the correlation function are required besides the scaling hypothesis.
Thereby, we take the first steps towards a numerical and theoretical study of the dynamics of quantum phase transitions and the QKZM for short-range interacting quantum matter in 2D. 
In a combined effort we employ a set of state-of-the-art numerical methods.
These include time evolution via a time-dependent variational principle (TDVP) for matrix product states (MPS)~\cite{Haegeman2016} or neural quantum states (NQS)~\cite{Carleo2017,ANN_Markus&Markus} on finite lattices, and 2D infinite projected entangled pair states (iPEPS) operating directly in the thermodynamic limit~\cite{CzarnikDziarmagaCorboz,ntu}.

Profiting from the individual strengths of each of the numerical approaches, we find strong evidence for universal dynamical behavior consistent with the scaling properties predicted by the QKZM.
We observe that the scaling regime is accessible already at moderate sweep rates and system sizes, 
in accord with the recent experiments on Rydberg atom quantum simulator platforms.
While we can identify the predicted QKZM scaling in the vicinity of the quantum phase transitions with high fidelity, we also observe deviations upon sweeping deeper into the ferromagnetic phase.
To account for it, we introduce an {\it extended quantum Kibble-Zurek mechanism} (xQKZM), which recognizes the effect of the modifications of the system's spectrum during the sweep and their influence on the excitation energies. 
%
%%%%%%%%%%%%%%%%%%%%%%%%%%%%%%%%%%%%%%%%%%%%%%%%%%%%%%%%%%%%%%%%%%%%%%%%%%%%%%
\begin{figure}[t]
\begin{centering}
\includegraphics[width=\columnwidth]{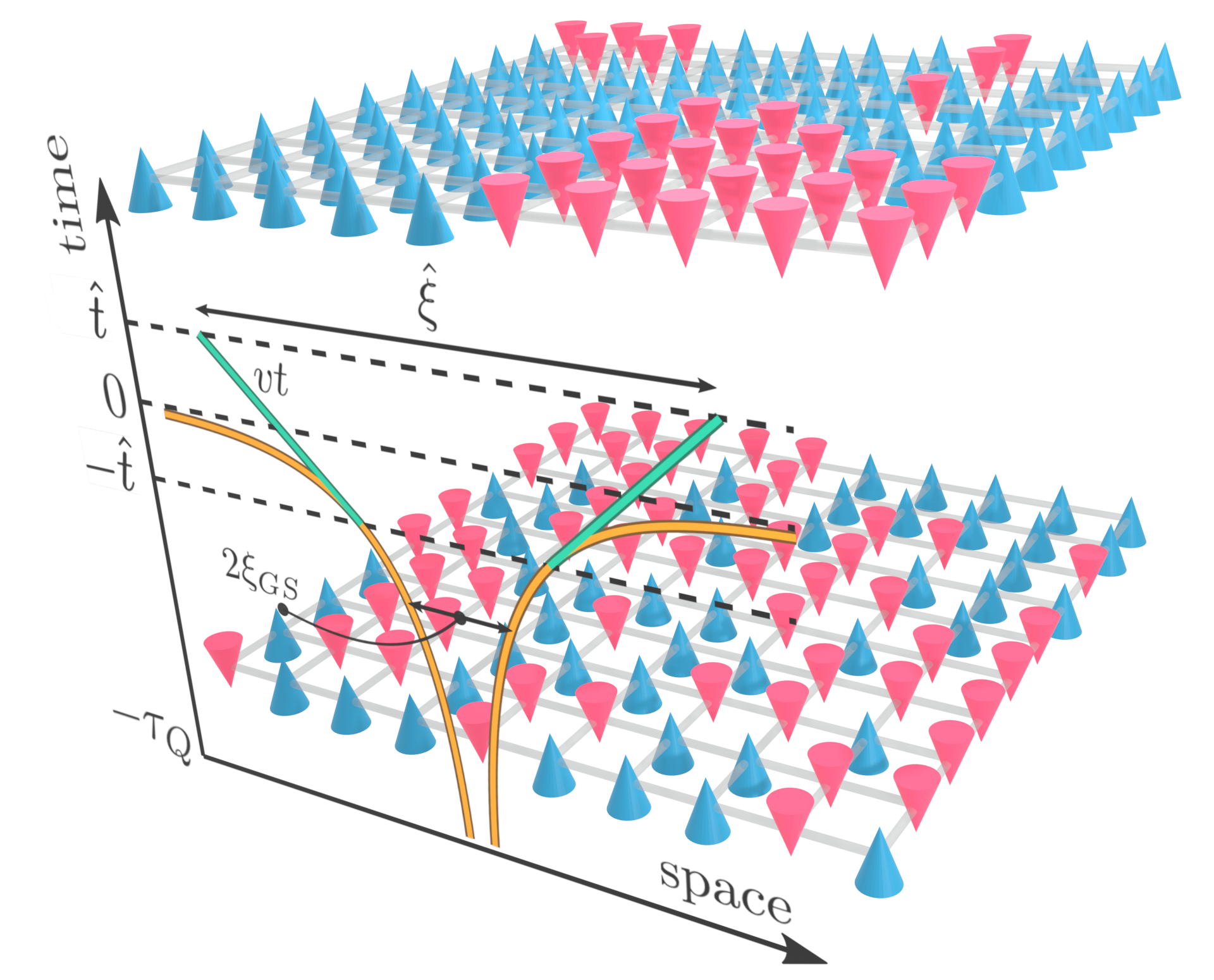}
\par\end{centering}
\caption{ 
{\bf Schematic depiction of the dynamics across a phase transition in a two-dimensional spin-$1/2$ model.}
In the initial paramagnetic state (bottom) spins align with the direction of the transverse magnetic field.
A measurement of the spin configuration in that state along the ordering direction would then typically yield a random pattern of spins pointing up (blue cones) or down (red cones). After a slow ramp across a quantum critical point the system develops a quantum superposition of ferromagnetic domains which upon measuring spin configurations along the ordering direction will yield typically a collapse onto a mosaic of such domains (top). On the front face, we include the growth of  the  ferromagnetic correlation range as a function of time $t$ starting from $t=-\tau_Q$ as the ramp progresses across the critical regime with the critical point located at $t=0$. The healing length $\hat\xi$ that determines the size of domains in the Kibble-Zurek mechanism is set at the characteristic time $|t| < \hat t$,  where the growth rate of the instantaneous ground state correlation length $\xi_{GS}$ exceeds the maximal speed of the relevant sound, $v$, in the system.
\label{fig:cartoon}}
\end{figure}
%%%%%%%%%%%%%%%%%%%%%%%%%%%%%%%%%%%%%%%%%%%%%%%%%%%%%%%%%%%%%%%%%%%%%%%%%%%%%%

\section{Results}
{\it Quantum KZ mechanism.---}
%\label{sec:QKZM}
%%%%%%%%%%%%%%%%%%%%%%%%%%%%%%%%%%%%%%%%%%%%%%%%%%%%%%%%%%%%%%%%%%%%%%%%%%%%%%
%
Quantum phase transitions occur in ground states of quantum many-body systems. They mark the singular points where quantum phases of matter transform into each other.
In the vicinity of these respective quantum critical points, the macroscopic physical properties become universal as a consequence of a divergent correlation length $\xi$. QKZM extends this universality -- it applies not only to static but also to dynamical properties. 

Of paradigmatic importance is the QKZM prediction of universal defect production upon quench though a quantum critical point.
Such dynamical crossing can be parametrized by the distance from a quantum phase transition through a dimensionless Hamiltonian parameter $\epsilon$.
Close to the critical point, the correlation length $\xi$ in the ground state diverges according to $\xi\propto|\epsilon|^{-\nu}$ and the energy gap closes according to $\Delta E\propto \xi^{-z}$, where $\nu$ and $z$ are the universal correlation length and dynamical critical exponents, respectively.

To study QKZM, a quantum Ising system is initially prepared in a paramagnetic ground state of its Hamiltonian.
It is subsequently smoothly ramped across a quantum critical point to the symmetry-broken state by varying that Hamiltonian.
Close to the critical point, the ramp can be linearized
\begin{equation}
\epsilon(t)=\frac{t}{\tau_Q},
\label{epsilont}
\end{equation}
with $t$ denoting the time. 
Consequently, $t=0$ corresponds to the quantum critical point in the fully adiabatic limit, and the quench time $\tau_Q$ sets the speed of the ramp.
As long as the evolution is adiabatic, the respective adiabatic correlation length, $\xi\propto|\epsilon|^{-\nu}$, increases at the rate:  
\be 
\frac{d\xi}{dt}=
\frac{d\epsilon}{dt}\frac{d\xi}{d\epsilon}\propto
\frac{1}{\tau_Q}
\frac{\nu}{|\epsilon|^{\nu+1}},
\ee 
which diverges at the critical point. However, perturbations and excitations of the order parameter in a quantum many-body system have a limited maximal speed of propagation $v$ (e.g., the speed of the relevant sound). Therefore, at some point, the actual speed at which correlation length can increase will not be able to keep up.
This results in the so-called {\it sonic horizon}. It determines the size of the domains that can choose broken symmetry in unison, as shown schematically in Fig.~\ref{fig:cartoon}.

Consequently, as the critical point is approached, there exists a time $-\hat t$ where the rate ${d\xi} / {dt}$ exceeds the relevant sounds speed $v$. For $z=1$ we have that $v=\mathrm{const.}$ whereas for $z\not=1$ the velocity becomes scale-dependent. 
For general $z$ we have that structures of size $\xi$ are subject to the relevant velocity $v \propto \xi^{-(z-1)} \propto \epsilon^{\nu(z-1)}$.
Comparing $d\xi/dt$ with $v$ results in a characteristic time scale
\be 
\hat t \propto\tau_Q^{z\nu/(1+z\nu)}
\label{hatt}
\ee 
at which the sonic horizon is determined.
The scaling obtained in this way is the same as in the adiabatic-impulse approximation~\cite{Z-a, *Z-b, *Z-c,sonic}. The corresponding healing length is set at $\hat t$ to be:
\begin{equation}
\hat\xi \propto \tau_Q^{\nu/(1+z\nu)}.
\label{hatxi}
\end{equation}
At large length and time scales, $\hat\xi$ and $\hat t$ specify the quench-induced evolution of the system near the critical point.

\begin{figure}[t!]
\begin{centering}
\includegraphics[width=\columnwidth]{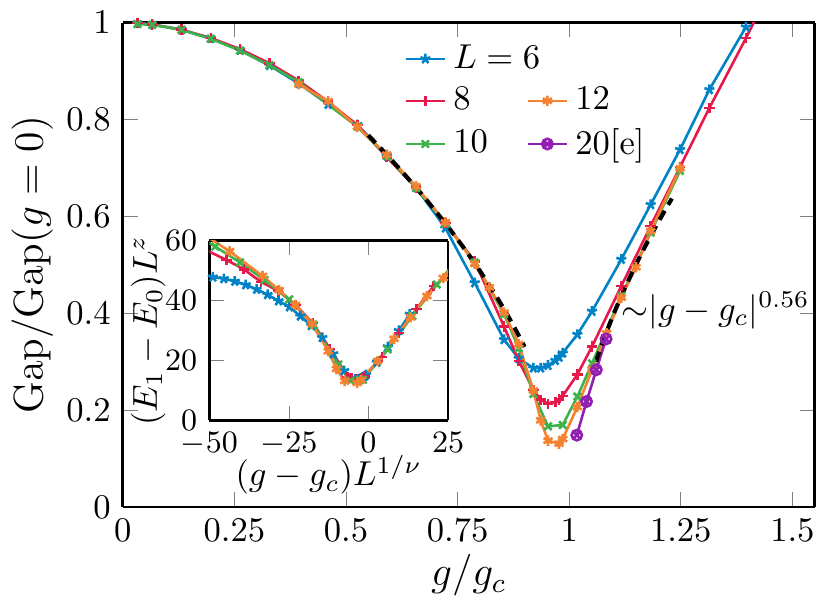}
\par\end{centering}
\caption{{\bf Energy gap as a function of the transverse field $g$ in a periodic lattice.} 
We collect results for different linear system sizes $L$ and fixed $J=J_c=1$. The data was obtained using the NQS approach for excited states, see the main text and Methods.
The inset shows the collapse of the data after finite size rescaling with the known critical exponents $z=1$ and $\nu\approx0.63$.
The collapse on the paramagnetic side was used to extrapolate the gap to $L=20$ in the main panel.
The black dashed lines on both sides of the transition -- in the regimes near estimated $\pm\hat t$, where the data are also converged in $L$ -- are consistent with exponent $z\nu=0.56$ that is shifted by a small non-universal $7\%$ correction from the exact $0.63$.
\label{fig:gap}}
\end{figure}

%%%%%%%%%%%%%%%%%%%%%%%%%%%%%%%%%%%%%%%%%%%%%%%%%%%%%%%%%%%%%%%%%%%%%%%%%%%%%%
{\it Setting and methods.---}
%%%%%%%%%%%%%%%%%%%%%%%%%%%%%%%%%%%%%%%%%%%%%%%%%%%%%%%%%%%%%%%%%%%%%%%%%%%%%%
Motivated by the recent experiments in Rydberg atom quantum simulators~\cite{rydberg2d1,rydberg2d2} and by its paradigmatic theoretical relevance, we consider in the following the 2D quantum transverse-field Ising Hamiltonian on a square lattice:
\begin{equation}
H(t) = - J(t) \sum_{\langle m,n\rangle} \sigma^z_m \sigma^z_n - g(t) \sum_{m=1}^{L^2} \sigma^x_m.
\label{eq:hamiltonian}
\end{equation}
Here, $\sigma_l^{\alpha}$, $\alpha=x,y,z$, denotes the Pauli matrices on lattice site $l$ with the linear extent of the system, $L$, implying overall $L^2$ lattice sites.
The model exhibits a quantum phase transition at $g_c/J_c= 3.04438$~\cite{Deng_QIshc_02}.
In Fig.~\ref{fig:gap}, we show results for the energy gap between the ground state and the first excited state in the zero momentum and even parity sector as a function of the transverse-field strength $g$ for periodic boundary conditions obtained using NQS and the approach to obtain excited states introduced in Ref.~\cite{Choo2018}.
In the vicinity of the quantum phase transition, as expected, one can observe significant finite-size effects. 
Nonetheless, the collapse of the data after finite-size rescaling with the known critical exponents $z=1$ and $\nu=0.629971$, which we show in the inset, reveals consistency with the expected universal behavior.
%

%%%%%%%%%%%%%%%%%%%%%%%%%%%%%%%%%%%%%%%%%%%%%%%%%%%%%%%%%%%%%%%%%%%%%%%%%%%%%%
%%%%%%%%%%%%%%%%%%%%%%%%%%%%%%%%%%%%%%%%%%%%%%%%%%%%%%%%%%%%%%%%%%%%%%%%%%%%%%

%
In the analysis of the time-evolved system, however, we will later see that the best collapse of the data is achieved assuming $z\nu=0.56$ to be the value of the product of the two critical exponents. The dashed lines in Fig.~\ref{fig:gap} show power laws of this form fitted to the data on both sides of the transition in regimes where finite size effects are small. These fits show that the data is on these finite intervals consistent with $z\nu\approx0.56$. We attribute this to sub-leading corrections which, for the considered ramp times $\tau_Q$ that we can numerically achieve, still yield a noticeable contribution, and whose influence can be effectively captured by slightly modified critical exponents.

Of central importance is a quantitative estimate of prefactor in the time scale $\hat t$, whose general scaling form has already been presented in Eq.~\eqref{hatt}.
For what follows, we define $\hat t_{\rm actual}$ as the time at which the rate $|\dot\epsilon(\hat t)/\epsilon(\hat t)|=|\hat t^{-1}|$ approximately equals the gap $\Delta E\propto(g_c\epsilon)^{z\nu}$ (setting the prefactor in that equality to one). With our fitted value for the prefactor of the gap opening on the paramegnetic side, we obtain
\begin{align}
    \hat t_{\rm actual} \approx 
    \left(\frac{\tau_Q^{z\nu}}{
    %9.6(4)g_c^{z\nu}
    11.6(3)
    }\right)^{\frac{1}{1+z\nu}} =
    %0.157(4)\tau_Q^{0.36(2)}
    0.208(4)\tau_Q^{0.36} \, .
    \label{hatt_fitted}
\end{align}
The main uncertainty in the fit originates from varying the fitting range. 
The domain of the fitted curve on the paramagnetic side in Fig.~\ref{fig:gap} corresponds to the gap at $\hat t$ for $0.8\leq\tau_Q\leq6.4$. 
Notice that the value of the prefactor in Eq.~\eqref{hatt_fitted} depends on the choice of an arbitrary $\mathcal O(1)$ prefactor when equating the rate with the gap.

To study the QKZM in our model, we initialize the system in the ground state $|\psi(t=t_i)\rangle = |{\rightarrow\rightarrow\rightarrow\ldots\rightarrow}\rangle$ of the Hamiltonian~\eqref{eq:hamiltonian} for $J/g=0$, where all spins align along the transverse field.
We then solve numerically the Schr\"odinger equation $i\partial_t |\psi(t) \rangle = H(t) |\psi(t)\rangle$. We fix the unit of time by setting $J_c=1$ (and $\hbar=1$).
Throughout this work, we study different sweep protocols in order to ensure that our observations are independent of the protocol details. On the one hand, we consider a {\it linear} quench:
\begin{eqnarray}
g(t)/g_c = 1-\epsilon(t),~~~ J(t)/J_c = 1+\epsilon(t), 
\label{eq:ramp_linear} 
\end{eqnarray}  
with $\epsilon(t)$ following Eq.~\eqref{epsilont}, starting at $t_i=-\tau_Q$ with $J(t_i)=0$, crossing the critical point at $t=0$, and ending at $t_f = \tau_Q$. As this sweep exhibits a nonanalytic temporal behavior at the starting $t_i$, and therefore might potentially generate further excitations masking the targeted QKZ features, 
we complement our analysis also by a {\it smooth} ramp, where the dimensionless distance follows
\begin{equation}
\tilde \epsilon(t) = \frac{t}{\tau_Q} - \frac{4}{27} \frac{t^3}{\tau_Q^3},
\label{eq:ramp_smooth} 
\end{equation}
between $t_i=-\frac{3}{2}\tau_Q$ and $t_f = \frac{3}{2}\tau_Q$. It has a vanishing first time-derivative at $t_i$ limiting the generation of additional excitations at the start of the protocol. Both ramps exhibit the same slope in the vicinity of the quantum critical point around $t\approx 0$, which, according to the general QKZM argument, is expected to result in identical universal scaling (we show in the Supplementary Material that those additional excitations are negligible compared to QKZM excitations).

\begin{table}[t!]
\center
\begin{tabular}{l|c|c|c|c}
Method& boundary cond. & $L$ & $\tau_Q$ & $t_f$ \\\hline
iPEPS& n.a. & $\infty$ &$\lesssim3.2$& $\lesssim 2~\hat t_{\rm actual}$ \\
NQS& periodic & $\lesssim 20$ &$\lesssim \tau_Q^{\rm adiab}$& $\lesssim \hat t_{\rm actual}$ \\
MPS & open & $\lesssim 14 $& any &$\tau_Q$
\end{tabular}
\caption{The employed numerical methods have complementary strengths. The table summarizes their rough ranges of applicability. Here $L$ is the linear lattice size and $\hat t_{\rm actual}$ is the value of $\hat t$ in Eq.~\eqref{hatt_fitted} with approximated prefactor, and $\tau_Q^{\rm adiab}$ estimates the crossover to adiabatic transition at $\hat \xi / L \approx 0.2$, see Fig.~\ref{fig:finite}.}
\label{tab:method_overview}
\end{table}

For the simulation of the quantum many-body dynamics, we employ three different numerical techniques, allowing us, on the one hand, to perform mutual cross-checks and, on the other hand, to cover complementary regimes of applicability. For a summary of the individual strengths and limitations of the methods, see Table~\ref{tab:method_overview}. First, we consider iPEPS as a tensor network method that operates directly in the thermodynamic limit of a two-dimensional system. Notably, our iPEPS simulations have been done with the neighborhood tensor update code \cite{ntu}, which provides a more stable upgrade of the full update code~\cite{CzarnikDziarmagaCorboz}. The second method is based on MPS, for which a one-dimensional ordering of the lattice sites is chosen to represent the wave function. Thereby we can simulate finite systems with open boundary conditions. The MPS wave function is evolved using the TDVP algorithm~\cite{Haegeman2016}. Finally, we use neural network quantum states, a recently proposed class of variational wave functions~\cite{Carleo2017}, to simulate finite systems with periodic boundary conditions. The time evolution was computed using convolutional neural networks and regularization techniques introduced in Ref.~\cite{ANN_Markus&Markus}. Furthermore, we also use NQS wave functions to variationally obtain the first excited states in addition to the ground state in order to extract information about the energy gap we show in Fig.~\ref{fig:gap}.

%%%%%%%%%%%%%%%%%%%%%%%%%%%%%%%%%%%%%%%%%%%%%%%%%%%%%%%%%%%%%%%%%%%%%%%%%%%%%%
\begin{figure}[t!]
\begin{centering}
\includegraphics[width=1\columnwidth]{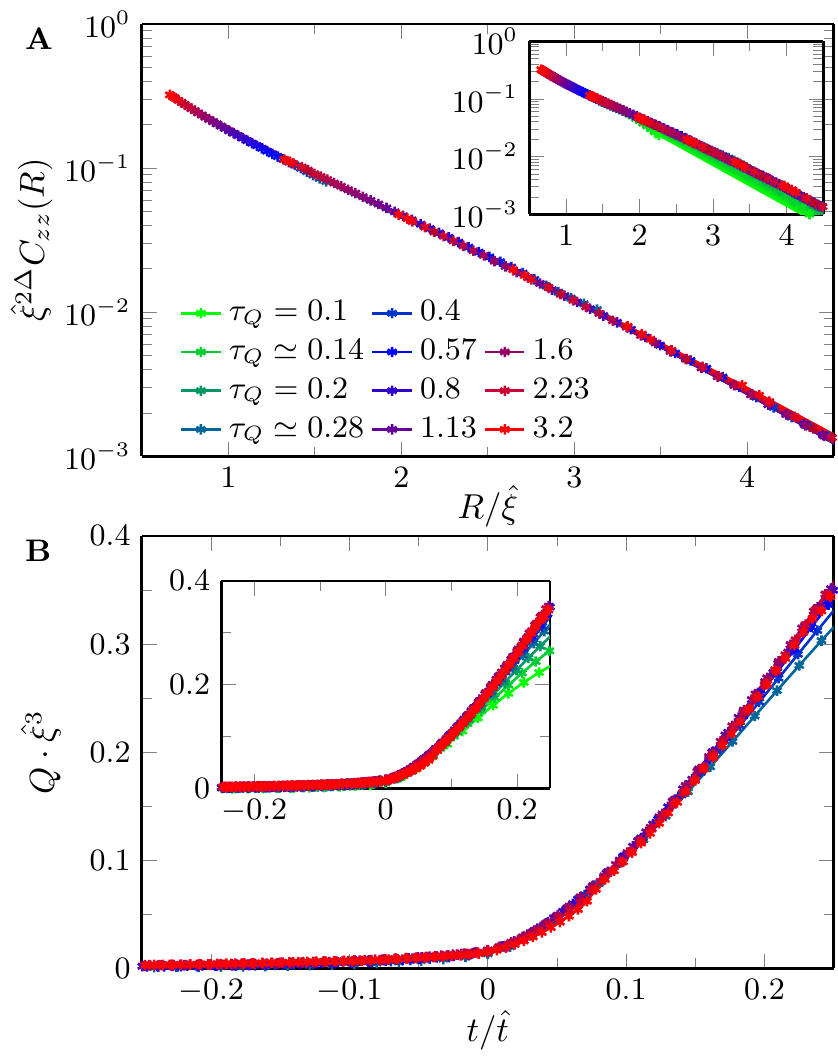}
\end{centering}
\caption{
\label{fig:ipeps}
{\bf Kibble-Zurek dynamical scaling in 2D quantum Ising model; infinite lattice.}
In {\bf A}, we collect the scaled ferromagnetic correlation function at the critical point, $\hat\xi^{2 \Delta} C^{zz}(t=0,R)$, as a function of the scaled distance, $R/\hat\xi$, and in {\bf B} a scaled excitation energy per site, $\hat\xi^{3} Q$, as a function of the scaled time, $t/\hat t$ (the critical value of the field is reached at $t=0$). The main panels show the collapse of the data for slower quenches with $\tau_Q\geq 0.28$, in agreement with the dynamical scaling hypothesis.  We obtain the best collapse for $\hat t=\hat\xi=\tau_Q^{0.36}$, where for rescaling we fix the prefactor in Eqs.~\eqref{hatt} and \eqref{hatxi} to one. The exponent that we obtain for the available (limited) $\tau_Q$'s is less than $10\%$ below the expected one of $0.386$.  Insets show a full range of quench times $\tau_Q=0.1\cdot2^{m/10}=0.1,...,3.2$, with integer $m$ in {\bf A}, and sparser data with integer $m/5$ in {\bf B}. 
}
\end{figure}
%%%%%%%%%%%%%%%%%%%%%%%%%%%%%%%%%%%%%%%%%%%%%%%%%%%%%%%%%%%%%%%%%%%%%%%%%%%%%%

%%%%%%%%%%%%%%%%%%%%%%%%%%%%%%%%%%%%%%%%%%%%%%%%%%%%%%%%%%%%%%%%%%%%%%%%%%%%%%
{\it Universal behavior in the thermodynamic limit.---}
%\label{sec:KZregime}
%%%%%%%%%%%%%%%%%%%%%%%%%%%%%%%%%%%%%%%%%%%%%%%%%%%%%%%%%%%%%%%%%%%%%%%%%%%%%%
We first consider the ramped quantum Ising model in the putative QKZ regime for times $-\hat t<t<+\hat t$ on an infinite lattice in the thermodynamic limit by means of iPEPS simulations.
We probe the system's properties mainly via two quantities.
The first one is the ferromagnetic correlation function
\begin{equation} \label{eq:Czz}
    C^{zz}(t,R) = \langle \psi(t) | \sigma_m^z \sigma_n^z | \psi(t) \rangle,
\end{equation}
where $R$ is the distance between the spins $m,n$.
Concretely, we compute $C^{zz}(t,R)$ for spins aligned along one of the axes.
Second, we consider the excitation energy density
\begin{equation}
    Q = \frac{1}{L^2} \left( \langle H(t)\rangle - E_0(t)  \right),
\end{equation}
where $\langle H(t) \rangle = \langle \psi(t) | \mathcal{H} | \psi(t) \rangle $ denotes the time-dependent expectation value of the Hamiltonian $H(t)$ at time $t$ with $|\psi(t)\rangle$ being the numerically exact solution of Schr\"odinger's equation.
Moreover, $E_0(t)$ is the ground state energy of $H(t)$ at parameter values $g(t)$ and $J(t)$.

In Fig.~\ref{fig:ipeps}{\bf A}, we show the correlation function $C^{zz}(t,R)$ at $t=0$ in units rescaled by the correlation length $\hat \xi$ for various ramp times $\tau_Q$.
It is a central result of our work that we observe a data collapse for $C^{zz}(t,R)$ upon utilizing the known scaling dimension $\Delta$ with $2\Delta=1+\eta$, where $\eta=0.036298(2)$~\cite{Deng_QIshc_02}.
Our data directly aligns with the QKZM prediction implying the following scaling form~\cite{KZscaling1,PhysRevB.85.100505,KZscaling2,Francuzetal}:
\be 
\hat\xi^{2 \Delta} C^{zz}(t,R) = F_C\left(t/\hat\xi^z,R/\hat\xi\right),
\label{CRscaling}
\ee
with $F_C$ a non-universal scaling function.
Overall, this prediction is expected to be asymptotically exact in a coarse-grained sense in the long-wavelength and low-frequency limit, corresponding to long ramp times $\tau_Q$.
In Fig.~\ref{fig:ipeps}{\bf A}, we find that the scaling regime can already be reached for rather small $\tau_Q$'s, which appears as a promising result from the experimental perspective.
Let us emphasize, however, that we obtain the best data collapse for the slower quenches in the range $0.3<\tau_Q<3.2$ for $\hat\xi\propto\tau_Q^{0.36}$.
The value $0.36$ for the exponent aligns directly with exponent identified for $\hat t$ using the energy gap, see Eq.~\eqref{hatt_fitted}.
As already discussed before, we see an error of about $7\%$ as compared to the asymptotically expected $z\nu/(1+z\nu)=0.3865$, which suggests that subleading corrections beyond the asymptotic universal behavior still yield a weak but noticeable contribution (see the Supplementary Material for a comparison of the qualities of collapse for the two values of the exponent).

In Fig.~\ref{fig:ipeps}{\bf B}, we quantify the number of defects as measured by the excitation energy density $Q$ in the putative scaling regime.
In the asymptotic limit of $\tau_Q\to\infty$, it is expected from the QKZM in $d$ spatial dimensions that the excitation energy density $Q$ also follows a universal behavior according to~\cite{KZscaling1,PhysRevB.85.100505,KZscaling2,Francuzetal}:
\begin{equation}
    \hat\xi^{d+z}~ Q = F_Q\left(t/\hat\xi^z\right).
    \label{scalingenergy}
\end{equation}
Here $d$ is a number of space dimensions. In Fig.~\ref{fig:ipeps}{\bf B}, we observe a data collapse of the properly scaled $Q$ in a time window in the vicinity of the critical point.
As for the ferromagnetic correlation function, the best collapse for the slower quenches is obtained for  $\hat\xi\propto\tau_Q^{0.36}$, also consistent with Eq.~\eqref{hatt_fitted}.

%%%%%%%%%%%%%%%%%%%%%%%%%%%%%%%%%%%%%%%%%%%%%%%%%%%%%%%%%%%%%%%%%%%%%%%%%%%%%%
\begin{figure}[t!]
\begin{centering}
\includegraphics[width=1\columnwidth]{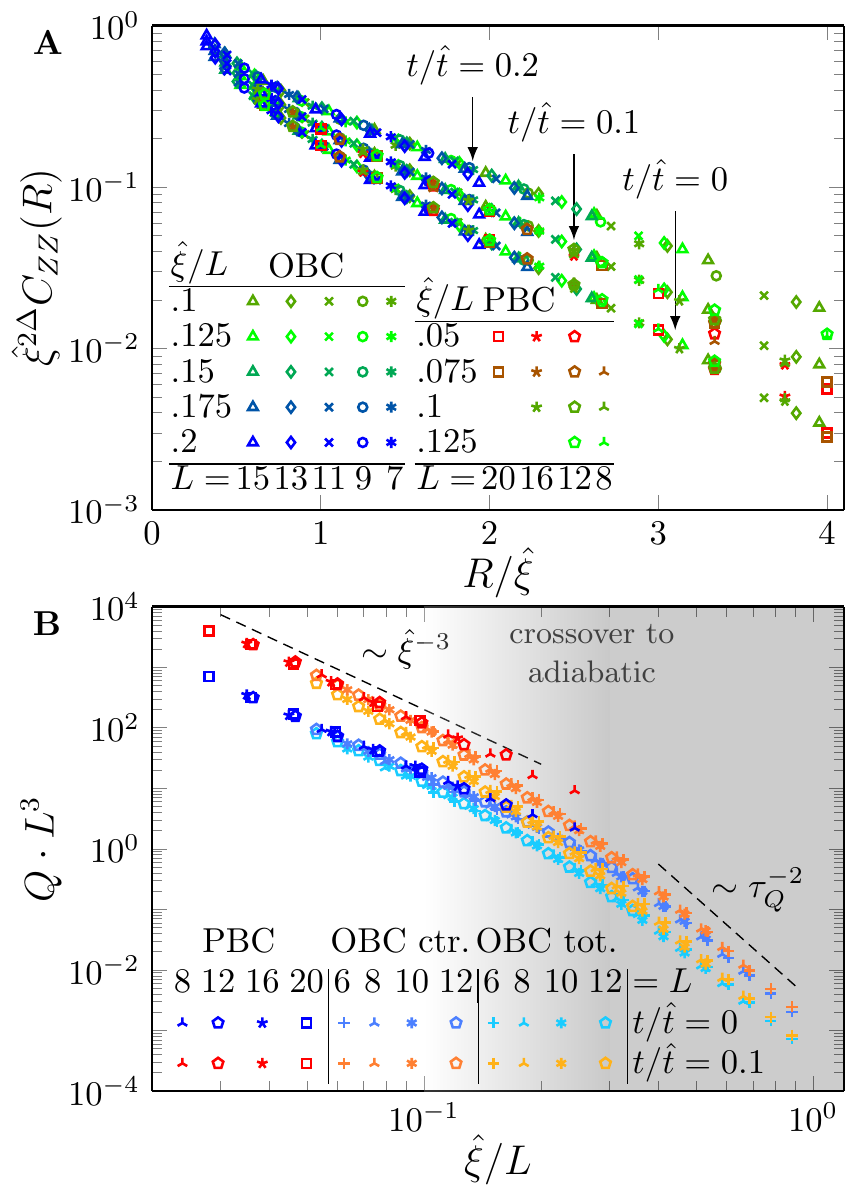}
\vspace{-5pt}
\end{centering}
\caption{
{\bf Kibble-Zurek dynamical scaling in 2D quantum Ising model; finite-size scaling.} We show the data collapse corroborating the dynamical scaling hypothesis, combining the results for open and periodic boundary conditions. In {\bf A}, we show the correlation function scaled according to Eq.~\eqref{CRscaling}, where finite-size effects are not appreciable in the presented range of   $\hat \xi/L$. For OBC, we calculate the correlation function with respect to the central spin. In {\bf B}, we show the scaled excitation energy density following Eq.~\eqref{ExiL}. For OBC, we compare the total energy (per spin) with the contribution to the excitation energy coming from the central spins. The latter closely follows the PBC results---extending them toward the adiabatic limit, marked by the change of power-law slope.  We can estimate the quench rate marking the transition to adiabatic limit $\tau_Q^{adiab}$,  based on crossover happening around   $\hat \xi/L \approx 0.2$.
In this plot, we use the exponent  $\hat \xi = \tau_Q^{0.36}$ with $\tau_Q\ge0.28$, as in Fig.~\ref{fig:ipeps}, that provides the best collapse also in this case.\label{fig:finite}}
\end{figure}
%%%%%%%%%%%%%%%%%%%%%%%%%%%%%%%%%%%%%%%%%%%%%%%%%%%%%%%%%%%%%%%%%%%%%%%%%%%%%%

%%%%%%%%%%%%%%%%%%%%%%%%%%%%%%%%%%%%%%%%%%%%%%%%%%%%%%%%%%%%%%%%%%%%%%%%%%%%%%
\begin{figure*}[t!]
\begin{centering}
\includegraphics[width=2\columnwidth]{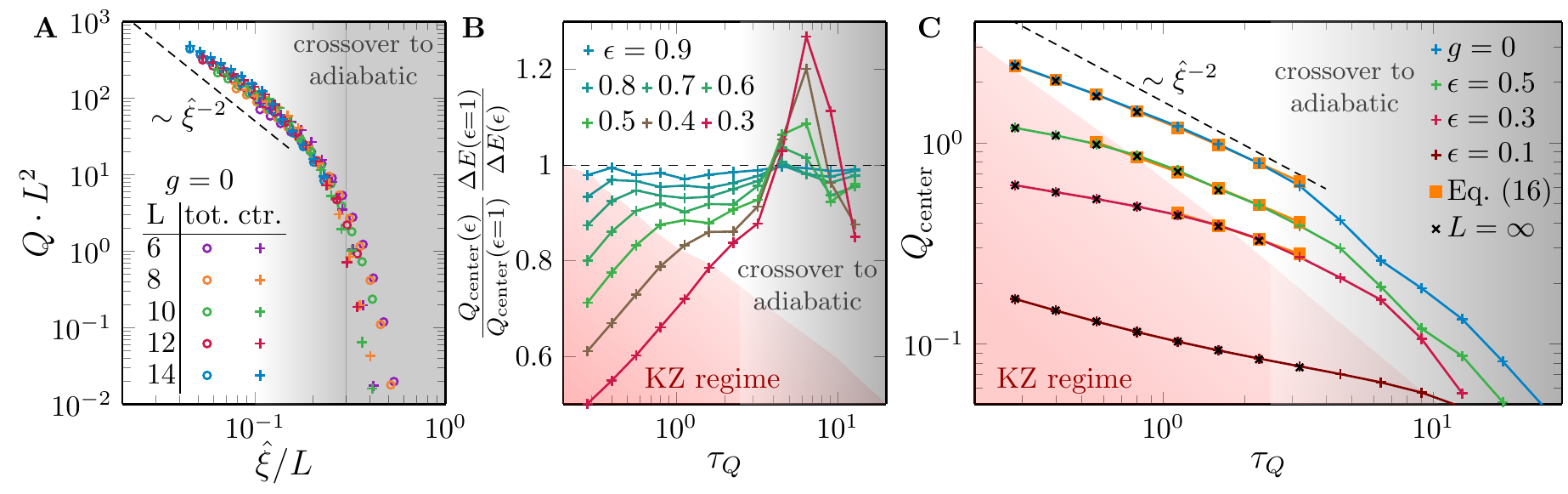}
\par\end{centering}
\caption{
{\bf Excitation energy measured deep in the ferromagnetic phase.} 
In {\bf A}, we plot the excitation energy per site at the final $g=0$ for a system with OBC. The data for different system sizes are plotted as a function of $\hat\xi/L=\tau_Q^{0.36}/L$, and the energy is scaled consistently with the relation $Q \propto \hat \xi^{-2} \propto \tau_Q^{-0.72}$ that follows from \eqref{eq:str}. This results in a relatively good -- but imperfect -- collapse for available system sizes. We show both the total energy (per spin) and the contribution coming from the central spins.
In {\bf B}, we test the conjecture of adiabatic excitation energy rescaling in Eq. \eqref{eq:str}, plotting the ratio of the excitation energy to the energy gap at different values of $\epsilon$, comparing it with the final $\epsilon=1$. The flattening of the curves in the central part of the panel supports the conjecture. Here, we use $L=14$ with OBC.
Finally, in {\bf C}, we quantify the deviation from the pure power-law dependence on $\tau_Q$ that follows from the simple conjecture in Eq.~\eqref{eq:str}. Snapshots of excitation energies at other values of $g$ show its build-up with an increasing energy gap (decreasing $g$). Here we calculate energy contribution for central spins in a system with OBC and $L=14$, which, in the KZ-regime, we also compare with the available iPEPS results (crosses). We fit Eq.~\eqref{eq:str_log} with parameters $(a,b)$ equal to $(1.4, 0.16)$ for $g=0$ ($\epsilon=1$), $(1.2, 0.2)$ for $\epsilon=0.5$, and $(1.0, 0.3)$ for $\epsilon = 0.3$. The fit is restricted to the regime of validity of Eq.~\eqref{eq:str_log} in between the KZ regime and the adiabatic regime. We use $\hat \epsilon$ calculated at twice the estimate in Eq.~\eqref{hatt_fitted} --  that is also used to single out part of dynamics within the universal KZ-regime (red filled area).
\label{fig:g0}}
\end{figure*}
%%%%%%%%%%%%%%%%%%%%%%%%%%%%%%%%%%%%%%%%%%%%%%%%%%%%%%%%%%%%%%%%%%%%%%%%%%%%%%

%%%%%%%%%%%%%%%%%%%%%%%%%%%%%%%%%%%%%%%%%%%%%%%%%%%%%%%%%%%%%%%%%%%%%%%%%%%%%%
{\it Universal behavior in a finite system.---}
%\label{sec:KZregime}
%%%%%%%%%%%%%%%%%%%%%%%%%%%%%%%%%%%%%%%%%%%%%%%%%%%%%%%%%%%%%%%%%%%%%%%%%%%%%%
After having explored the case of thermodynamically large systems, in the next step, we now address the experimentally relevant regime of large but finite system sizes with linear extent $L$.
In order to obtain a comprehensive picture, including the behavior both for open and periodic boundary conditions, we base our analysis on results obtained using MPS and NQS wave functions, where for MPS (open boundary conditions) we have a smooth ramp in Eq.~\eqref{eq:ramp_smooth}, and for NQS (periodic boundary conditions) we have a linear ramp in Eq.~\eqref{epsilont}.
The key consequence of considering a finite system is that the energy gap does not close in the vicinity of the critical point, as illustrated by our numerical results in Fig.~\ref{fig:gap}.
This implies that for sufficiently large $\tau_Q$, an additional adiabatic regime emerges, where the system asymptotically follows the ground state.
This finite-size effect can provide a further test of a generalized KZ scaling. On a general level, the respective crossover from QKZ to adiabatic scaling occurs when $\hat\xi \propto L$.
While for $\hat \xi \ll L$ the system follows the QKZ paradigm, the adiabatic regime is recovered for $\hat \xi \to L$.

In Fig.~\ref{fig:finite}{\bf A}, we focus first on the QKZ regime by showing results for the ferromagnetic correlation function $C^{zz}(t,R)$ for various $\tau_Q$ resulting in different $0.05 \leq \hat \xi/L \leq 0.2$. 
Here, we again observe a convincing data collapse irrespective of boundary conditions and quench details (smooth or linear), in line with the considerations outlined before, that $\hat \xi/L \ll 1$ is expected to yield the QKZ regime.
We also note that we identify a collapse of similar quality for times $-\hat t < t < \hat t$ within the predicted scaling regime.

Concerning the excitation energy density, we find power-law behavior for $\hat \xi \ll L$ consistent with
\begin{equation}
    Q\propto\hat\xi^{-3} \, ,
\end{equation}
which follows directly from the general QKZM prediction that the excitation energy density is supposed to exhibit the following scaling form~\cite{KZscaling1,PhysRevB.85.100505,KZscaling2,Francuzetal}:
\begin{equation}
    \hat\xi^{d+z} Q = F_Q\left(t/\hat\xi^z,\hat\xi/L\right).
    \label{ExiL}
\end{equation}
In particular, we emphasize that for $t=0$, our numerical finding of algebraic $\xi^{-3}$-dependence extends over more than one decade.
For larger $\tau_Q$ and consequently upon approaching $\hat \xi \to L$ , we also observe the expected deviations towards the adiabatic regime, where $Q \propto \tau_Q^{-2}$.
Let us note that the finite-size gap differs for open and periodic boundary conditions implying that also the crossover scale between QKZ and adiabatic is slightly shifted with respect to each other.

%%%%%%%%%%%%%%%%%%%%%%%%%%%%%%%%%%%%%%%%%%%%%%%%%%%%%%%%%%%%%%%%%%%%%%%%%%%%%%
{\it Extended quantum Kibble-Zurek mechanism.---}
%\label{sec:g0}
%%%%%%%%%%%%%%%%%%%%%%%%%%%%%%%%%%%%%%%%%%%%%%%%%%%%%%%%%%%%%%%%%%%%%%%%%%%%%%
After exploring the QKZ scaling in the vicinity of the quantum critical point, we now take the next step by continuing the parameter ramp deep into the ferromagnetic phase down to $g=0$. 
Unlike short-range 1D systems, the 2D quantum Ising model supports a symmetry-broken phase at nonzero temperatures which can lead to new kinds of dynamics such as coarsening or phase-ordering kinetics.
These simulations, which correspond to long evolution times, turned out to be numerically the most challenging, and we could not fully converge the iPEPS and NQS simulations for this purpose. Therefore, we rely mostly on the MPS technique in the following.

In Fig.~\ref{fig:g0}{\bf A}, we display our numerical results for the excitation energy per site $Q$ as a function of $\tau_Q$ in rescaled units for finite system sizes obtained using MPS.
While we observe a kind of data collapse, as one might expect from the general picture of the QKZM, there does not appear a clear power-law behavior as the collapsed data exhibits a slight bending in the utilized double-logarithmic plot (apart from the expected crossover to the exponential scaling in the adiabatic limit at large $\tau_Q$).

\begin{figure}[t]
    \centering
    \includegraphics[width=\columnwidth]{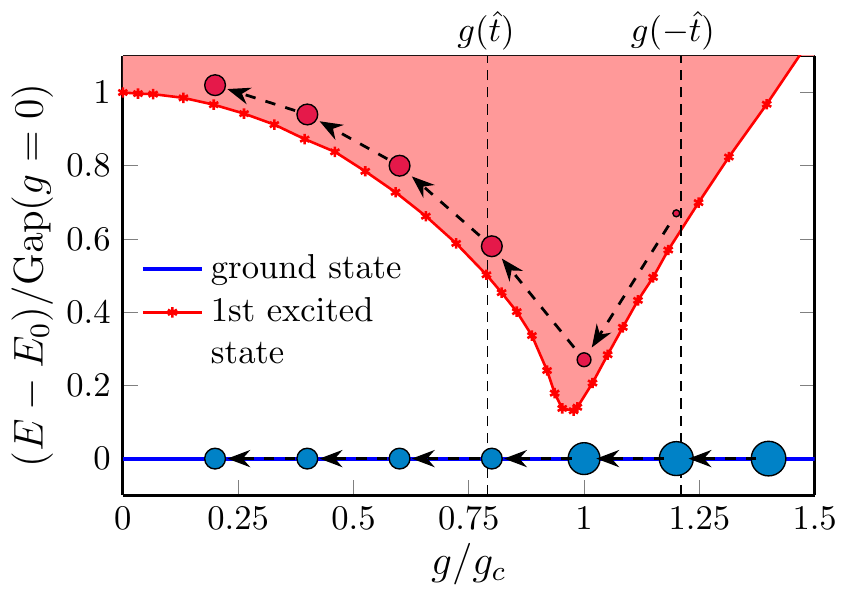}
    \caption{{\bf Schematic depiction of the xQKZM.} The lines in the plot show the energies of the ground state and of the first excited state (combined data from Fig.~\ref{fig:gap}); the shaded area represents further excited states. The circles indicate the occupation of the ground state and of the low-lying excited states. The assumption of the xQKZM is that for a given rate $\tau_Q$ excited state occupation is generated during the time interval $[-\hat t,\hat t]$. Subsequently, the occupations are assumed to be time-independent, but excitation energy is increased due to adiabatic gap rescaling. }
    \label{fig:xqkzm}
\end{figure}

To understand and quantify the deviations from the expected power-law dependence, we now formulate a simple conjecture to predict the excitation energy at the end of the ramp at $g=0$ far beyond the universal regime, which terminates around $t=\hat t$.
This conjecture we call the \textit{extended quantum Kibble-Zurek mechanism} (xQKZM), generalizing a concept established in 1D~\cite{Francuzetal} to our interacting 2D setting.
After $\hat t$ the evolution becomes adiabatic, i.e., the gap becomes large enough to prevent any transfer of occupation between the instantaneous ground state and excited states. 
The xQKZM is based on two assumptions, whose validity and limitations for the considered parameter regimes will be discussed later on: i) after $\hat t$ the redistribution of occupations among the instantaneous excited eigenstates can be neglected, so that energy changes can only emerge from the parametric dependence of the energy eigenvalues corresponding to the eigenstates during the sweep; ii) the details of the parametric dependence of the relevant energy eigenvalues can also be neglected so that the eigenvalues exhibit roughly a global rescaling by a scale set by the gap $\Delta E(\epsilon)$.
A schematic depiction displaying the parametric dependence of the gap and the occupations is shown in Fig.~\ref{fig:xqkzm}.
This yields the following estimate for the excitation energy $Q$ at the end of the ramp:
\be 
Q
\propto
\hat \xi^{-(d+z)}
\frac{\Delta E(\epsilon)}{\Delta E(\hat\epsilon)}
\propto
\hat\xi^{-d} \Delta E(\epsilon) .
\label{eq:str}
\ee 
In the first step we assume that the excitation energy at $+\hat\epsilon$, corresponding to the time $t=\hat{t}$ is proportional to $\hat \xi^{-(d+z)}$ in accordance with the scaling hypothesis in Eq.~\eqref{scalingenergy} and the data collapse in Fig.~\ref{fig:ipeps}{\bf B}.
In the second step, we assume a further simplification in that the gap scales as $\Delta E(\hat\epsilon)\propto\hat\epsilon^{z\nu}$ over the full range of considered $\hat\epsilon$; this assumption holds for small enough $\hat\epsilon$ (or, equivalently, slow enough quenches) and large enough system sizes such that the slowest quenches are still not adiabatic. In this regime, the simple conjecture predicts that in particular, for $g=0$, the excitation energy should scale as $\hat\xi^{-d}$. This power law is indicated by dashed lines in Fig.~\ref{fig:g0}{\bf A} and {\bf C}. 
While it captures the leading trend, our data show a deviation beyond the leading power-law behavior that we discuss later. 
An equation similar to \eqref{eq:str} appeared in Ref. \onlinecite{DeGrandi2010} but for small $\epsilon$ where $\Delta E(\epsilon)\propto\epsilon^{z\nu}$.

In Fig.~\ref{fig:g0}{\bf B} we test the xQKZM prediction including the ratio $\Delta E(\epsilon=1)/\Delta E(\epsilon)$ numerically by comparing the relative excitation energy change $Q(\epsilon=1)/Q(\epsilon)$ to the relative change $\Delta E(\epsilon=1)/\Delta E(\epsilon)$ of the gap between some intermediate values $\epsilon$ of the ramp and the end $\epsilon=1$. In accordance with Eq. \eqref{eq:str} we can see flattening of the curves in the central part of Fig.~\ref{fig:g0}{\bf B} that lies between the KZ regime on the left (where $\epsilon<\hat\epsilon$) and the adiabatic regime on the right. 

Let us now discuss the regime of validity of the proposed xQKZM conjecture.
Concerning the first assumption of neglecting the redistribution of occupations among eigenstates, this clearly depends on the overall considered quench times.
For sufficiently large $\tau_Q$ the general expectation would be that the system is supposed to undergo phase-ordering kinetics and coarsening dynamics, which would involve a second type of universal dynamical process and which originates precisely from redistribution of occupations.
However, for the $\tau_Q$ considered in the numerical computations of this work we are operating in a different regime.
While the $\tau_Q$ are still sufficiently large in order to observe numerical evidence for the QKZM, the overall time span of the dynamics in the ferromagnetic phase for times $t>\hat t$ is limited so that such redistribution can be for now approximately neglected.
Concerning the second assumption of a roughly uniform shift of the relevant energy eigenvalues deep in the ferromagnetic phase the respective validity depends crucially on whether the dominant occupation of the instantaneous eigenstates originates from states of the order of the instantaneous gap, or more specifically from eigenstates not too far up in the excitation spectrum.
Although the QKZM describes the creation of occupations in excited states, it is still reasonable to assume that these excitations are not dominantly located in the non-universal high-energy regime, where universality would anyway be out of reach.
Further, by increasing $\tau_Q$ the likelihood of generating high-energy excitations can be systematically decreased, so that the assumption ii) is more likely to be approximately valid.

In Fig.~\ref{fig:g0}{\bf C} we test the conjecture in Eq.~\eqref{eq:str} against our numerical data.
We find that the bare xQKZM already accounts for the main contributions to the excitation energies. Consistent with Fig.~\ref{fig:g0}{\bf B}, however, we also observe that a quantitative comparison requires to take into account corrections.
Empirically, we find that these are consistent with a logarithmic dependence on $\tau_Q$:
\begin{equation}
    Q (\epsilon) \simeq
Q(\hat \epsilon) 
\frac{\Delta E(\epsilon)}{\Delta E(\hat\epsilon)}(a + b \log(\tau_Q))
\label{eq:str_log}
\end{equation}
with $a,b$ some constants.
In Fig.~\ref{fig:g0} {\bf C} we include the xQKZM in combination with these logarithmic corrections to the numerical MPS data and observe a close correspondence.
The main influence of the logarithmic correction is to impose a bending of the excitation energy towards smaller $\tau_Q$.
Let us note that for the xQKZM data, we focus just on the bulk behavior to avoid boundary contributions, which are significant for the considered system sizes but are irrelevant in the thermodynamic limit.
For that purpose, we only show the MPS data for the energy in the center of the 2D lattice.
Further, assuming that the bulk gap is not affected by boundary conditions, we use our numerically obtained gap $\Delta E(\hat\epsilon)$ from Fig.~\ref{fig:gap} and the exact value $\Delta E(g=0)=16J$.
%

\begin{figure}[t!]
\begin{centering}
\includegraphics[width=\columnwidth]{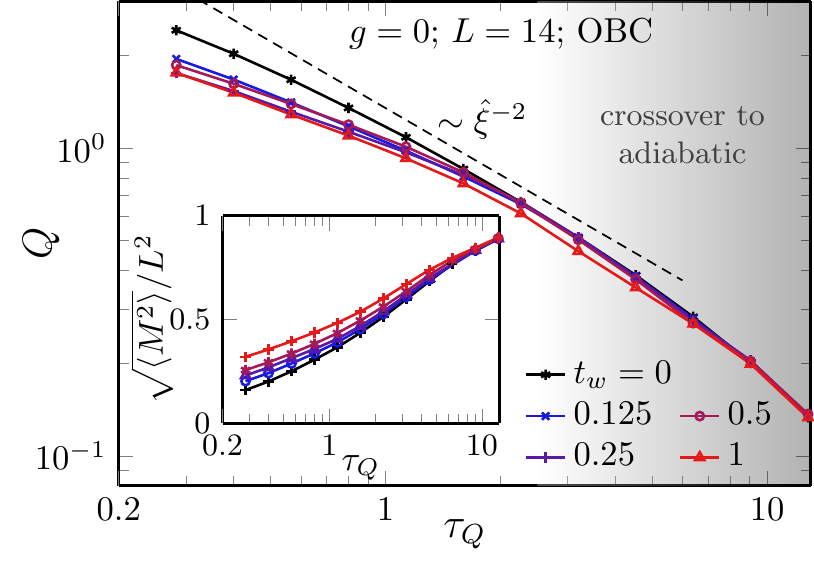}
\par\end{centering}
\caption{{\bf Influence of free evolution in the ferromagnetic phase on the final state at $g=0$.}
Here, we consider a protocol as in Fig.~\ref{fig:g0}, but with the ramp that stops at $\epsilon = 0.5$ deep in the ferromagnetic phase for a waiting time $t_w$. The evolution then continues to $g=0$ where the observables are measured. In the main panel, we show the total excitation energy per spin and, in the inset, correlations in the system measured by the variance of ferromagnetic magnetization. The energy decreases and quickly saturates with increasing $t_w$. In contrast, no such saturation is seen in the variance, where longer $t_w$ provides more time for magnetic ordering.
\label{fig:waiting}}
\end{figure}

%%%%%%%%%%%%%%%%%%%%%%%%%%%%%%%%%%%%%%%%%%%%%%%%%%%%%%%%%%%%%%%%%%%%%%%%%%%%%%
{\it Beyond xQKZM.---}
%\label{sec:g0}
%%%%%%%%%%%%%%%%%%%%%%%%%%%%%%%%%%%%%%%%%%%%%%%%%%%%%%%%%%%%%%%%%%%%%%%%%%%%%%
%
A central assumption in the xQKZM is that redistribution of occupations among eigenstates can be neglected.
This naturally neglects all scattering processes leading to thermalization or phase ordering kinetics which might be especially relevant in the 2D context studied here.
For that purpose, we study a slight variant of the dynamical protocol, which allows us to obtain some understanding of the influence of thermalization and phase ordering kinetics onto the excitation energy density $Q$.
Specifically, we interrupt the ramp when the transverse-field strength $g_s$ reaches a value $g_s/g_c=1/2$ for a waiting time $t_w$, where all Hamiltonian parameters are held constant, before continuing down towards $g=0$.
In this way, we provide further time for the system to relax and to redistribute occupations. 
The effect of the additional evolution on the outcome at $g=0$ is shown in Fig.~\ref{fig:waiting}, where we include both the final excitation energy density $\epsilon$ and the final magnetization fluctuations $\langle M ^2 \rangle = \sum_R C^{zz}(t,R)$ in the inset.
We can see that the final excitation energy density tends to decrease with respect to $t_w=0$, implying a kind of cooling effect due to the intermediate waiting interval.
Further, the magnetization fluctuations increase with $t_w$ consistent with coarsening, i.e., the tendency of the system to develop ferromagnetic order at sufficiently low transverse fields.
We attribute this observed path dependence to thermalization dynamics and phase ordering kinetics, which is not present in the 1D version of the model that is effectively noninteracting. It, however, becomes immediately relevant for the 2D case.
One can understand the observed energy decrease by first deriving the equation of motion for the internal energy $E(t) = \langle H(t) \rangle$, which we will show here for the linear quench:
\begin{equation}
    \label{eq:eom_energy}
    \frac{d E(t)}{d t} = - \frac{1}{\tau_Q} \sum_{\langle m,n\rangle} \langle \sigma^z_m(t) \sigma^z_n(t)\rangle  + \frac{1}{\tau_Q} \sum_m \langle \sigma_m^x(t)\rangle \, .
\end{equation}
While the energy itself is the sum of the spin-spin interaction and the transverse-field term, the energy change during the dynamics is governed by their \emph{difference}.
Now, it is crucial to realize that the thermalization dynamics and the resulting coarsening in our 2D Ising model is exactly characterized by a redistribution of energy between these individual contributions.
It is a central consequence of the ramp starting on the paramagnetic side, that the transverse magnetization is enhanced compared to the instantaneous equilibrium state. The accompanying thermalization dynamics is characterized by a transfer of energy from the transverse field to the interaction term.
%
%%%%%%%%%%%%%%%%%%%%%%%%%%%%%%%%%%%%%%%%%%%%%%%%%%%%%%%%%%%%%%%%%%%%%%%%%%%%%%
%%%%%%%%%%%%%%%%%%%%%%%%%%%%%%%%%%%%%%%%%%%%%%%%%%%%%%%%%%%%%%%%%%%%%%%%%%%%%%
According to Eq.~\eqref{eq:eom_energy}, this, in turn, also implies a negative change in the energy leading to a cooling effect.
In other words, thermalization and coarsening dynamics, which naturally favor the interaction energy compared to the transverse-field contribution, directly affect the final energy at $g=0$. 
These considerations also provide a direct interpretation of the results we observe in Fig.~\ref{fig:waiting}.
While the intermediate interruption of the ramp protocol itself does not change the energy directly, it gives the system the additional time $t_w$ to thermalize towards the instantaneous equilibrium state leading to an enhanced interaction and reduced field energy.
However, this implies a larger right-hand side in the magnitude of Eq.~\eqref{eq:eom_energy} when the protocol is resumed after $t_w$ implying a stronger reduction of the total energy as compared to the case without stopping.
Overall, Eq.~\eqref{eq:eom_energy} highlights, in combination with thermalization and coarsening properties of genuinely interacting 2D models, that there can be a noticeable
energy change occurring during the parameter ramp in the ordered phase.
%

%%%%%%%%%%%%%%%%%%%%%%%%%%%%%%%%%%%%%%%%%%%%%%%%%%%%%%%%%%%%%%%%%%%%%%%%%%%%%%
\section{Discussion}
%\label{sec:conclusion}
%%%%%%%%%%%%%%%%%%%%%%%%%%%%%%%%%%%%%%%%%%%%%%%%%%%%%%%%%%%%%%%%%%%%%%%%%%%%%%
%
In this work we have studied the QKZM in a 2D transverse-field Ising model utilizing the combined effort of state-of-the-art numerical methods.
As a main result, we have found universal defect production in the vicinity of the quantum critical point.
For parameter sweeps deep into the ferromagnetic phase, we have introduced an extended QKZM (xQKZM), which accounts for additional spectral information for the prediction of the final excitation energy densities.

The exponent that yields the best scaling collapse consistently deviates by about $7\%$ from the precise values of the critical exponents that have been determined in previous numerical studies. We attribute this deviation to the fact that all simulations were limited by maximal feasible ramping rates $\tau_Q$ or system sizes $L$. These limitations constrain our numerical experiments to probe a regime where corrections to the asymptotic universal scaling laws are relevant. In this regime, the gap opening is still approximately described by a power law, but the best fitting exponent differs from the known value in the asymptotic limit. Hence, we expect better agreement of the dynamically observed critical exponents with results from studies in equilibrium for larger system sizes and slower ramping rates. Our results, among others, help to estimate the necessary parameters. For instance, in order to avoid the finite-size effects for $\tau_Q \simeq 20$, assuming $\hat \xi/L \le 0.1$ in Fig.~\ref{fig:finite}, would require the system of linear size $L=30$.

Motivated in part by the remarkable progress in Rydberg atom and superconducting qubit quantum simulators, the theoretical analysis provided in this work exhibits a natural implementation in an experimental context.
While evidence for QKZM has already been observed in systems of Rydberg atoms for a one-dimensional quantum spin chain~\cite{Lukin18}, very recent developments enable the realization of the dynamics of transverse-field Ising models in two-dimensional geometries involving hundreds of spin degrees of freedom~\cite{rydberg2d1,rydberg2d2}.
Notice that although these systems typically realize antiferromagnetic spin interactions, the resulting dynamics is equivalent to that of an Ising ferromagnet upon transforming $\sigma_l^z \mapsto -\sigma_l^z$ on every other lattice site.
It is straightforward to implement the fully-polarized initial condition~\cite{2017Zeiher,Lukin18,rydberg2d1,rydberg2d2} and to temporally tune the couplings in order to realize the parameter sweep across the underlying quantum critical point~\cite{Lukin18,rydberg2d1,rydberg2d2}.
Further, measurements in these systems yield spin configurations along a tunable axis, e.g., along $\sigma^z$ or $\sigma^x$, which can give access to both the spin-spin correlation functions we considered, see Eq.~\eqref{eq:Czz}, and the total Ising energy, see Eq.~\eqref{eq:hamiltonian}.
Although quantum-optical systems such as Rydberg atoms exhibit remarkable isolation from the environment, decoherence nevertheless limits the experimentally accessible time scales.
In this context, it is important to emphasize that the recent experiments have demonstrated already long coherence times, such as to observe the QKZM in one dimension~\cite{Lukin18} and close-to-adiabatic preparation of symmetry-broken low-energy states~\cite{rydberg2d1,rydberg2d2}.
Overall, this makes our theoretical results directly experimentally accessible in Rydberg atom systems with the potential to address central questions that have remained open after the pioneering experiment on 2D QKZM in Ref. \onlinecite{rydberg2d1}.
This concerns for instance the QKZ scaling deep in the ferromagnetic regime for larger system sizes and longer ramp times as well as immediate evidence for scaling behavior illustrated by a data collapse of the full correlation function.

%%%%%%%%%%%%%%%%%%%%%%%%%%%%%%%%%%%%%%%%%%%%%%%%%%%%%%%%%%%%%%%%%%%%%%%%%%%%%%
\section*{Methods}
%%%%%%%%%%%%%%%%%%%%%%%%%%%%%%%%%%%%%%%%%%%%%%%%%%%%%%%%%%%%%%%%%%%%%%%%%%%%%%

In this work, we use a combination of three state-of-the-art numerical algorithms. They provide mutual cross-checks where their ranges of applicability overlap and their full range is broader than any individual one.

The first, in order of appearance, is the 2D iPEPS tensor network on an infinite lattice. To simulate time evolution, we used the neighborhood tensor update~\onlinecite{ntu} which is a more efficient version of the code in Ref. \onlinecite{CzarnikDziarmagaCorboz}. Here we use second-order Suzuki-Trotter decomposition with time step $dt=0.01$. Accuracy is limited in a controlled way by a bond dimension $D$ of the iPEPS ansatz. All presented results appear converged for $D=8$. Evaluation of expectation values requires approximating the infinite tensor environment whose accuracy is limited by an environmental bond dimension $\chi$. The results for $D=8$ were converged with $\chi\leq 32$. The iPEPS simulations fail after $t\approx 2\hat t_{\rm actual}$ where the convergence in $D$ becomes insufficient. 

The second method is based on representing the system's wave function as a one-dimensional matrix product state, where the 1D chain spans a 2D lattice. We obtain the best convergence (with respect to the required MPS bond-dimensions $D$) using a diagonal steps-like covering. The time evolution is simulated using the TDVP algorithm of Ref.~\onlinecite{Haegeman2016} (combining the one-site scheme with local application of two-sites updates for enlarging bond dimensions) and $4$th order time-dependent Suzuki-Trotter decomposition. We typically find the time-step $dt=1/8$ to be sufficient (with a few smaller time-steps at the beginning to avoid instability of the TDVP when applied to the initial product state). We simulate the system sizes up to $L=14$, with the bond dimension up to $D=512$ to converge the presented results in various limits. 

The third numerical approach employs neural quantum states (NQS) as variational ansatz for the wave function~\cite{Carleo2017}. This approach was recently shown to be suited for the accurate simulation of quench dynamics in the two-dimensional quantum Ising model~\cite{ANN_Markus&Markus}. The precision of NQS simulations is determined by the size of the neural network, which allows for systematic convergence checks. The time evolution was simulated using convolutional neural networks (CNNs) and regularization techniques introduced in Ref.~\cite{ANN_Markus&Markus}. For system sizes $L\leq10$, we used a single-layer network with fully connected filters. For larger systems, we employed two-layer networks with square filters that have a diameter of $L/2$. We checked different network sizes for convergence and found that 6 channels, or 4 followed by 3 channels, were sufficient in the single and two-layer case, respectively.

With the NQS, we can simulate finite systems with periodic boundary conditions in 2D, explicitly enforcing the invariance of the wave function under all lattice symmetries. This allows us to estimate boundary effects by comparing to MPS simulations, and the large system sizes reached convincingly reveal the power-law scaling of energy at the critical point. However, we found that the accuracy of the NQS simulations breaks down when continuing the ramp too far into the ferromagnetic phase; in particular, $g=0$ is currently out of reach.

To compute the energy gap, we implemented the algorithm for excited state search introduced in Ref.~\cite{Choo2018}. The first step is to perform the ground-state search using the stochastic reconfiguration algorithm as already established in~\cite{Carleo2017}. In the second step, the stochastic reconfiguration is modified such that the trial wave function is explicitly projected onto the subspace orthogonal to the ground state. We used the rescaled energy variance $\sigma_E=\braket{H^2-\braket{H}^2}/L^4$ to assert the accuracy of our result and reached for both the ground state and the first excited state in all cases $\sigma_E<10^{-4}$ and often $\sigma_E\approx10^{-5}$ within 250 optimization steps of size $\delta\tau=0.01$.

%%%%%%%%%%%%%%%%%%%%%%%%%%%%%%%%%%%%%%%%%%%%%%%%%%%%%%%%%%%%%%%%%%%%%%%%%%%%% 
%\acknowledgements
%%%%%%%%%%%%%%%%%%%%%%%%%%%%%%%%%%%%%%%%%%%%%%%%%%%%%%%%%%%%%%%%%%%%%%%%%%%%%
%
\noindent {\bf Acknowledgements}
We acknowledge funding by National Science Centre (NCN), Poland under projects 2019/35/B/ST3/01028 (JD) and NCN together with European Union through QuantERA ERA NET program ~2017/25/Z/ST2/03028 (MMR),
and Department of Energy under the Los Alamos National Laboratory LDRD Program (WHZ).
WHZ was also supported by the U.S. Department of Energy, Office of Science, Basic Energy Sciences, Materials Sciences and Engineering Division, Condensed Matter Theory Program.
This project has received funding from the European Research Council (ERC) under the European Union’s Horizon 2020 research and innovation programme (grant agreement No. 853443), and M. H. further acknowledges support by the Deutsche Forschungsgemeinschaft via the Gottfried Wilhelm Leibniz Prize program.
MS was supported through the Leopoldina Fellowship Programme of the German National Academy of Sciences Leopoldina (LPDS 2018-07).
Parts of the numerical simulations were performed at the Max Planck Computing and Data Facility in Garching.
Moreover, the authors gratefully acknowledge the Gauss Centre for Supercomputing e.V. (www.gauss-centre.eu) for funding this project by providing computing time through the John von Neumann Institute for Computing (NIC) on the GCS Supercomputer JUWELS at J\"ulich Supercomputing Centre (JSC)~\cite{JUWELS}.

%%%%%%%%%%%%%%%%%%%%%%%%%%%%%%%%%%%%%%%%%%%%%%%%%%%%%%%%%%%%%%%%%%%%%%%%%%%%%%%

\noindent {\bf Data Availability} 
All data needed to evaluate the conclusions in the paper are present in the paper and/or the Supplementary Material.

\noindent {\bf Author Contributions} 
All the authors contributed equally to the preparation of the manuscript, discussions, and interpretation of the results.
Numerical simulations have been performed by M.S. (NQS), M.M.R. (MPS) and J.D. (iPEPS).

\noindent {\bf Additional information}
Correspondence and requests for materials should be addressed to M.S. (email:markus.schmitt@uni-koeln.de) or M.M.R. (email:marek.rams@uj.edu.pl).

\noindent {\bf Competing interests}
The authors declare no competing interests.

%%%%%%%%%%%%%%%%%%%%%%%%%%%%%%%%%%%%%%%%%%%%%%%%%%%%%%%%%%%%%%%%%%%%%%%%%%%%%%%%%%%%%%%%%
%\bibliographystyle{ScienceAdvances.bst}
\bibliography{KZref.bib}
%%%%%%%%%%%%%%%%%%%%%%%%%%%%%%%%%%%%%%%%%%%%%%%%%%%%%%%%%%%%%%%%%%%%%%%%%%%%%%%%%%%%%%%%%
%\clearpage
%\newpage
%%%%%%%%%%%%%%%%%%%%%%%%%%%%%%%%%%%%%%%%%%%%%%%%%%%%%%%%%%%%%%%%%%%%%%%%%%%%%%%%%%%
%\appendix
%%%%%%%%%%%%%%%%%%%%%%%%%%%%%%%%%%%%%%%%%%%%%%%%%%%%%%%%%%%%%%%%%%%%%%%%%%%%%%%%%%%



\section*{Supplementary material}

# Quantum phase transition dynamics in the two-dimensional transverse-field Ising model

Markus Schmitt
*Institute for Theoretical Physics, University of Cologne, 50937 Cologne, Germany*

Marek M. Rams and Jacek Dziarmaga
*Jagiellonian University, Institute of Theoretical Physics, Łojasiewicza 11, PL-30348 Kraków, Poland*

Markus Heyl
*Max Planck Institute for the Physics of Complex Systems, Nöthnitzer Straße 38, Dresden 01187, Germany*

Wojciech H. Zurek
*Theory Division, Los Alamos National Laboratory, Los Alamos, New Mexico 87545, USA*

(Dated: July 29, 2022)

## LINEAR VERSUS SMOOTH RAMP

In the main text we consider a straight linear ramp and one that is smoothed at the ends. The straight ramp with its sharp beginning (discontinuous time derivative) is inconvenient for tensor networks (TN) simulations because from the very start it generates some excitations that have nothing to do with the KZ mechanism in question but whose entanglement has to be accounted for by extra bond dimension of the networks. This does not seem to be a problem for the neural network (NN) and that is why the NN employs the linear ramp that also requires less time to simulate. For the same reason it would be more convenient to perform in an experiment.

The precise beginning and ending of the ramp does not affect the KZ mechanism because KZ excitations are set near the critical point which is right in the middle between the beginning and the end. This can be seen in e.g. Fig. 4**A** where there is no appreciable difference between NN and TN results. Small discrepancies between NN and TN in Fig. 4**B** are solely due to their different boundary conditions which are, respectively, periodic and open and, therefore, the discrepancies are finite size effects, compare Fig. S1 below where open boundary conditions are assumed for both the linear and the smooth ramp. This independence on ramp's beginning/ending is very convenient for experimental implementations giving them some flexibility.

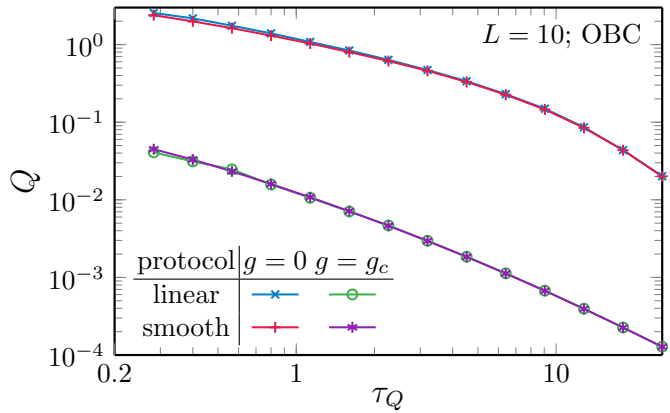

Figure S1. **Independence of the KZ mechanism on the ramp details.** We compare the excitation energy for a linear ramp in Eq. (1) of the main text, and a ramp that is smoothed at the ends as in Eq. (8) of the main text. Both ramps have the same slope – set by $\tau_Q$ – when crossing the critical point. We show excitation energy per spin measured at the critical point, and at the end of the ramp when $g = 0$. The details of the ramp have a marginal influence on the resulting excitation energy. There is a very small "cooling" effect related to a slightly longer time for coarsening dynamics in the smooth ramp that is visible for $g = 0$ and fast quenches – consistent with trends in Fig. 7. This is, however, a sub-leading effect.

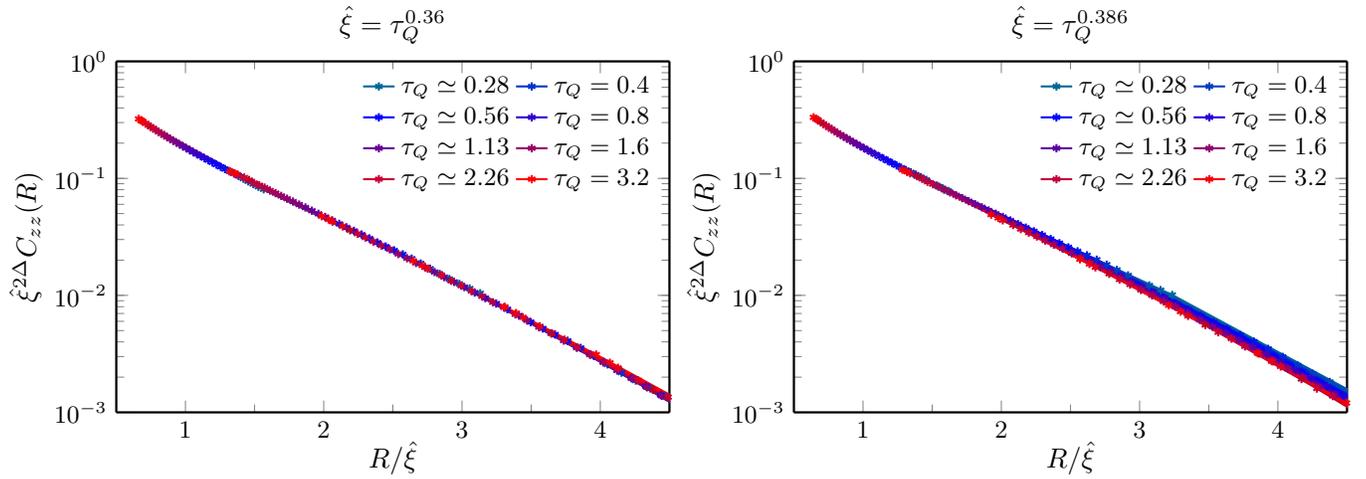

Figure S2. **The best collapse of the correlation function.** Scaled correlation function for 36 values of $\tau_Q = 0.1 \cdot 2^{m/10}$, with $m = 15, \ldots, 50$. In the left panel we assume $\hat{\xi} = \tau_Q^{0.36}$ and in the right one $\hat{\xi} = \tau_Q^{0.386}$. The collapse on the left is clearly better than on the right for available range of quench times $\tau_Q$.

## DYNAMICAL EXPONENT

In Fig. S2, we compare collapses of the correlation function obtained by iPEPS on an infinite lattice for the range of quench times $0.28 \leq \tau_Q \leq 3.2$ with the exponents $0.36$ and the exact $0.386$. Considering the longest $\tau_Q$ that we can achieve here, the effective $0.36$ clearly yields a better collapse than $0.386$. The collapse for $0.386$ is still decent, but mainly because the two exponents differ by just 7%. The precision of our numerical data is good enough to discriminate this small difference.

## EXTRAPOLATION OF THE ENERGY GAP

The inset in Fig. 2 of the main text tests the finite-size scaling hypothesis for the energy gap near the critical point. The collapse is compelling, especially on the paramagnetic side. In Fig. S3 below, we use the data on the paramagnetic side to extrapolate the gap in system size up to $L = 100$. The extrapolated data agrees with the expected power-law with exponent $z\nu \approx 0.63$, and give an impression of the significance of finite size effects close to the critical point. The extrapolation to $L = 20$, which is the size we use to simulate quenches, is also shown in Fig. 2 of the main text.

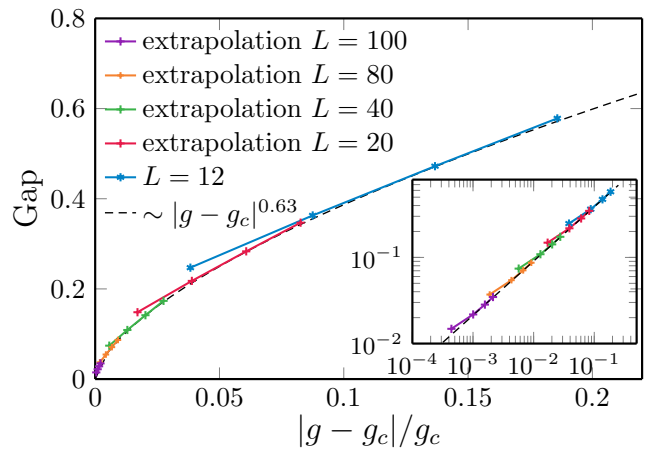

Figure S3. **Extrapolation of the energy gap in larger system sizes.** We extrapolate the energy gap on the paramagnetic side of the critical point for a few large system sizes based on the finite-size scaling collapse shown in the inset of Fig. 2 of the main text.



%%%%%%%%%%%%%%%%%%%%%%%%%%%%%%%%%%%%%%%%%%%%%%%%%%%%%%%%%%%%%%%%%%%%%%%%%%%%%%%%%%%%%%%%%
\end{document}